\documentclass[a4paper,11pt]{article}
\usepackage{amssymb,amsmath,amsfonts,placeins,pbox,multirow,mathtools}
\usepackage{graphicx,rotate,color,slashed,cite,epstopdf,verbatim,url,multirow,booktabs}
\usepackage{jheppub}
\usepackage{xcolor}
\usepackage{epstopdf}
\usepackage{subfig}
\usepackage{bm}
\usepackage{graphicx,caption}
\usepackage{array}
\usepackage{soul}
\usepackage[T1]{fontenc}

\numberwithin{equation}{section}
\newcommand{\tr}{\mathrm{tr}}

\def\ssbh#1//#2//{\ensuremath{\xrightarrow [\substack{#2}]
    {\parbox{3cm}{\hfil $\scriptstyle \langle #1 \rangle$ \hfil}}}}

\usepackage[capitalise]{cleveref}
\crefname{section}{Sec.}{Secs.}
\crefname{table}{Tab.}{Tabs.}
\crefname{figure}{Fig.}{Figs.}
\crefname{equation}{Eq.}{Eqs.}
\crefname{appendix}{Appendix}{Appendix}

\title{$\!\!\!\!$One-loop Renormalization Group Equations in Generic$\!\!\!\!$ Effective Field Theories. Part I: Bosonic Operators.}

\author{Miko{\l}aj Misiak}
\author{and Ignacy Nałęcz}

\emailAdd{misiak@fuw.edu.pl}
\emailAdd{inalecz@fuw.edu.pl}

\affiliation{Institute of Theoretical Physics, Faculty of Physics, University of Warsaw,\\
ul. Pasteura 5, 02-093 Warsaw, Poland}

\date{} 
\abstract{
We consider a generic class of effective quantum field theories with
arbitrary gauge groups and scalar matter fields. In such theories, we
derive the one-loop Renormalization Group Equations (RGEs) for the
physical dimension-six operators. The present paper is the first one
in a series that is going to cover theories with spin-$\frac12$ matter
fields, too, including the phenomenologically most relevant SMEFT and
LEFT cases. Our present approach provides tools for deriving the yet
unknown two-loop RGEs in the SMEFT.
}

\begin{document}
\maketitle

\section{Introduction}
\label{sec:intro}

Quantum Field Theories (QFTs) provide a framework for describing
physical phenomena across a wide range of energy scales. It often
happens that masses of certain particles are much larger than the
characteristic energy scale of the considered process. In such a case,
the well-known decoupling procedure can be applied, and we pass to the
Effective Field Theory (EFT) formalism. This approach has been
successful in different branches of particle physics. A notable
application is the Low-energy Effective Field Theory (LEFT), derived
from the Standard Model (SM) by decoupling the massive vector bosons,
the top quark, and the Higgs boson. The LEFT has been successfully applied
to derive highly accurate predictions for branching ratios and
particle lifetimes in weak decays, surpassing the precision of direct
calculations in the full SM~\cite{Buchalla:1995vs}. Another important
application of the EFTs is the Standard Model Effective Field Theory
(SMEFT), often employed in the search for beyond-SM physics. Within
the SMEFT paradigm, the SM is treated as a low-energy EFT emerging
from a yet unknown more fundamental theory, after decoupling of some
heavy degrees of freedom~\cite{Weinberg:1980wa,Buchmuller:1985jz,Grzadkowski:2010es}.

The EFT degrees of freedom are the light fields that have not been
decoupled, contrary to the heavy ones. In the Lagrangian, one must
include local operators of arbitrary dimensionality that are built of
the light fields and their derivatives. The structure of such
operators is restricted by the gauge and global symmetries of the
theory. In many phenomenological applications, it is sufficient to
consider operators up to dimension five and six, as the experimental
precision does not require higher accuracy. The Wilson coefficients
(WCs) of such operators are determined through matching with the
original theory. In the matching procedure, the Green's functions of
the light particles in the EFT are compared to those in the original
full theory (or a higher-level EFT). The WCs are fixed by ensuring
that the EFT reproduces the full theory results to the desired order
of the expansion in $m/\Lambda$, where $m$ stands for the light
particle masses and momentum products, while $\Lambda$ is the lightest
decoupled particle mass. If the matching calculation is performed in
dimensional regularization, then the renormalization scale is set to
be of order $\Lambda$, which ensures absence of large logarithms that
would worsen the perturbation series behaviour in the WC
determination. To use the EFT at lower energy scales, one has to find
the WCs at lower renormalization scales. For this purpose, one derives
and solves the Renormalization Group Equations (RGEs) for the WCs.

In this paper, we derive one-loop RGEs for the physical dimension-six
operators in a wide class of EFTs. We focus on the operators involving
only the scalar and gauge fields, calling them ``bosonic
operators''. An extension of our analysis to theories involving
spin-$\frac12$ matter fields is postponed to a future
publication~\cite{Mieszkalski:2026}. As far as the gauge group is
concerned, we allow it to be any finite product of Lie groups whose
algebras belong to the classical set $\{A_n,B_n,C_n,D_n\}$.

The leading terms in the EFT Lagrangian consist of operators of
dimension smaller or equal four. In their case, the relevant couplings
for which the RGEs are derived are not called the WCs but rather the
gauge couplings, masses, Yukawa couplings, as well as the triple- and
quartic scalar couplings. Although the WCs of higher-dimensional
operators do affect their renormalization group evolution, such effects are usually
neglected, as they generate only ${\mathcal O}(m/\Lambda)$ corrections to
observables that do not vanish in the $\Lambda \to \infty$ limit. On
the other hand, we do encounter observables that vanish (or are very
small) in this limit, and they are the most important in the EFT
applications. Therefore, in the case of couplings at dimension $\leq
4$ operators, we content ourselves with the one- and two-loop RGEs
that were derived for generic renormalizable QFTs in
Refs.~\cite{Machacek:1983tz,Machacek:1983fi,Machacek:1984zw}, taking
into account more recent verifications and corrections of these
results~\cite{Luo:2002ti,Schienbein:2018fsw}. At the one-loop level,
we confirm the findings of Ref.~\cite{Schienbein:2018fsw}.  Beyond two
loops, generic RGEs for the gauge and Yukawa couplings can be
found in Refs.~\cite{Pickering:2001aq,Poole:2019kcm,Bednyakov:2021qxa,Davies:2021mnc}.

Recently, a classification of dimension-six operators alongside with
some preliminary results for the RGEs of bosonic dimension-six
operators was presented by us in section (2.8) of
Ref.~\cite{Aebischer:2023nnv}, following the master
theses~\cite{Nalecz:2021xx,Mieszkalski:2021xx}. However, no complete
RGEs in the physical (on-shell) basis have been published to date,\footnote{
See, however, the ``note added'' at the end of section~\ref{sec:summ}.}
even for the operators containing the bosonic fields only.

To simplify our calculations, we assume a discrete symmetry $\phi \to
-\phi$, which removes odd-dimensional operators but has no effect on
even-dimensional ones thanks to Lorentz invariance. For more general
EFTs without such a discrete symmetry, the RGEs for odd-dimensional
operators can still be obtained by treating one of the scalar fields
as an auxiliary gauge-singlet with fixed vacuum expectation value.
The same method is, in principle, applicable for future
determination of dimension-five operator effects on the dimension-six
operator RGEs. However, this would require extending our calculation
to diagrams with double operator insertions. In the SMEFT and LEFT
cases, such effects of dimension-five operators are negligible in most
practical applications either due to smallness of the neutrino masses
(SMEFT), or to chirality suppression of the dipole operators
(LEFT).

To derive the necessary Feynman rules, we take advantage of the
background field method~\cite{Abbott:1980hw}. In consequence, all the
counterterms we need to consider at one loop originate from gauge
invariant operators that contain no quantum gauge field but rather the
classical background gauge field only. Within this framework, it is
actually advantageous to begin with deriving the RGEs in an off-shell
basis, i.e.\ including the unphysical operators that vanish by the
Equations Of Motion (EOM). Only in the next step the
basis is reduced to the on-shell one, via applying the EOM.\footnote{
More precisely, via performing appropriate field
redefinitions~\cite{Criado:2018sdb} that are equivalent to applying
the EOM in the context of one-loop RGEs.}
While there is certainly some additional work associated with such an
approach, it does provide non-trivial checks of our results, as the
final RGEs in the on-shell basis must depend neither on the reducible
operator WCs, nor on the gauge-fixing parameter.

Our paper is organized as follows. In section~\ref{sec:class}, we
define the considered class of theories and determine a particular
off-shell basis of dimension-six operators.  Next, in
section~\ref{sec:On-shell}, we derive the transformation to the
on-shell basis. In section~\ref{sec:RGE}, we describe our methodology
and present our main result, namely the one-loop RGEs for all the
physical bosonic operators. We summarize our findings in
section~\ref{sec:summ}, leaving a few technical details to the appendix.

\section{Classification of dimension-six operators}
\label{sec:class}

\subsection{The model definition and notational conventions}

In a general relativistic EFT with gauge and scalar
degrees of freedom, all the matter fields can be written in
terms of real ones $\phi_a$. After the $\phi \to -\phi$ symmetry
has been imposed, the Lagrangian of such a theory takes the
form
\begin{align}\label{eq:Lagrangian}
{\mathcal L} &= -\frac14 F^A_{\mu\nu} F^{A\,\mu\nu} + \frac12 (D_\mu\phi)_a (D^\mu\phi)_a
- \frac12 m^2_{ab} \phi_a\phi_b -\frac{1}{4!} \lambda_{abcd} \phi_a \phi_b \phi_c \phi_d\nonumber\\
&\quad + {\mathcal L}_{\rm g.f.} + {\mathcal L}_{\rm FP}+ \frac{1}{\Lambda^2} \sum Q_N + {\mathcal O}\left(\frac{1}{\Lambda^4}\right), 
\end{align}
where $Q_N$ denote linear combinations of dimension-six operators
multiplied by their WCs, and
\begin{align*}
&F^A_{\mu\nu}  = \partial_\mu V^A_\nu - \partial_\nu V^A_\mu - f^{ABC} V^B_\mu V^C_\nu,\\
&(D_\rho F_{\mu\nu})^A  = \partial_\rho F^A_{\mu\nu} - f^{ABC} V_\rho^B F^C_{\mu\nu},\\
&(D_\mu\phi)_a = \left(\delta_{ab}\partial_\mu + i\theta^A_{ab} V^A_\mu\right) \phi_b.
\end{align*}
Here, the gauge couplings have been absorbed into the definitions
of structure constants and matter field representation generators.  In
Eq.~\eqref{eq:Lagrangian}, ${\mathcal L}_{\rm g.f.}$ and ${\mathcal
L}_{\rm FP}$ stand for the gauge-fixing and Fadeev-Popov terms, while
${\mathcal O}\left(\frac{1}{\Lambda^4}\right)$ denotes higher-order
operators suppressed by powers of the $\Lambda$ scale beyond which the
EFT is not valid any more, and the full theory must be used
instead.

The generators $\theta^A$ are hermitian and antisymmetric, which
implies that all their components are imaginary. They satisfy the
usual relation $[\theta^A,\theta^B] = i f^{ABC} \theta^C$. For any
non-abelian gauge boson $V^{\underline B}$, we can select the
corresponding simple subgroup of the overall gauge group. Let us
denote it by $G_{\underline B}$, and its (possibly reducible)
representation in the scalar field space by $S_{\underline B}$. We
work in a basis where $f^{{\underline B}DE}f^{CDE} =
C_2(G_{\underline{B}}) \delta^{\underline{B}C}$. Using the Schur's
lemma for the (irreducible) adjoint representation of $G_{\underline
B}$, one easily proves that for any two generators $\theta^A$ and
$\theta^C$ of $G_{\underline B}$, the trace $\;\tr\!\left(\theta^A
\theta^C\right)$ is proportional to $\delta^{AC}$. The proportionality
constant will be denoted by $S_2(S_{\underline B})$, i.e.
\begin{equation} \label{eq:trace}
\tr\!\left(\theta^A \theta^C \right) = S_2(S_{\underline B}) \delta^{AC}.
\end{equation}
%
%
%
%
As far as the possible $U(1)$ (or, equivalently, $SO(2)$) gauge fields
are concerned, we choose their basis in such a way that no kinetic
mixing arises at the tree level, and the product of any two different
$U(1)$ generators in the scalar-field representation is
traceless. This is always possible, as for any two quadratic forms in
$\mathbb{R}^n$, one of them being positive-definite, there exists a
basis where both forms have diagonal matrices. In such a basis, no
$U(1)$ gauge bosons mix under one-loop renormalization, and
Eq.~(\ref{eq:trace}) holds for both the abelian and non-abelian
generators.

\subsection{Dimension-six operator basis}

We renormalize the dimension-six terms in our Lagrangian~\eqref{eq:Lagrangian}
using the background-field formalism~\cite{Abbott:1980hw}.  The renormalization is first
completed in the off-shell basis that contains all the dimension-six
terms allowed by the symmetries. The off-shell basis reads\footnote{
See section~(2.8) of Ref.~\cite{Aebischer:2023nnv} for an extension of
our off-shell basis to theories involving spin-$\frac12$ fermionic
fields.}
\begin{align}\label{eq:Basis_off_shell}
Q_1    &= \frac{1}{6!} W^{( 1)}_{abcdef}\, \phi_a \phi_b \phi_c \phi_d \phi_e \phi_f,&
Q_2    &= \frac{1}{4} W^{( 2)}_{abcd}\, (D_\mu\phi)_a (D^\mu\phi)_b \phi_c \phi_d,\nonumber\\
Q_3    &= \frac12 W^{( 3)}_{ab}\, (D^\mu D_\mu\phi)_a (D^\nu D_\nu\phi)_b,&
Q_4    &= \frac12 W^{( 4)A}_{ab}\, (D^\mu\phi)_a (D^\nu\phi)_b F^A_{\mu\nu},\nonumber\\
Q_5    &= \frac14 W^{( 5)AB}_{ab}\, \phi_a \phi_b F^A_{\mu\nu}F^{B\,\mu\nu},&
Q_6    &= \frac14 W^{( 6)AB}_{ab}\, \phi_a \phi_b F^A_{\mu\nu}\widetilde F^{B\,\mu\nu},\nonumber\\
Q_7    &= \frac12 W^{( 7)AB}\, \left(D^\mu F_{\mu\nu}\right)^A \left( D_\rho F^{\rho\nu}\right)^B,&
Q_8    &= \frac{1}{3!} W^{( 8)ABC}\, F^{A\,\mu}_{~~~\;\nu}F^{B\,\nu}_{~~~\;\rho}F^{C\,\rho}_{~~~\;\mu},\nonumber\\
Q_9    &= \frac{1}{3!} W^{( 9)ABC}\, F^{A\,\mu}_{~~~\;\nu}F^{B\,\nu}_{~~~\;\rho}\widetilde F^{C\,\rho}_{~~~\;\mu},&
\end{align}
where the real tensors $W^{(N)}$ contain both the WCs and the
necessary Clebsch-Gordan coefficients that select singlets from
various tensor products of the gauge group representations.  In
general, each $W^{(N)}$ contains many independent WCs, and many
gauge-singlet operators are present in each $Q_N$. We emphasize that
only some components of the tensorial $W$-coefficients are
independent, as the operator structure imposes various symmetries on
the $W$-coefficients. The complete list of such symmetries is as
follows:
$W^{(1)}_{abcdef} = W^{(1)}_{(abcdef)}$, 
$W^{(2)}_{abcd}   = W^{(2)}_{(ab)(cd)}$,
$W^{(3)}_{ab}     = W^{(3)}_{(ab)}$,
$W^{(4)A}_{ab}    = W^{(4)A}_{[ab]}$,
$W^{(5)AB}_{ab}   = W^{(5)(AB)}_{(ab)}$,
$W^{(6)AB}_{ab}   = W^{(6)(AB)}_{(ab)}$,
$W^{(7)AB}        = W^{(7)(AB)}$,
$W^{(8)ABC}       = W^{(8)[ABC]}$,
$W^{(9)ABC}       = W^{(9)[ABC]}$,
where the round or square brackets enclosing the indices denote their
symmetrization or antisymmetrization, respectively.

In addition, gauge invariance of the theory imposes 
essential identities on the couplings and $W$-coefficients. To
derive such identities, one considers infinitesimal
transformations of the fields in a given operator. For
instance, for the scalar mass term
\begin{equation}
\begin{aligned}
m^2_{ab}\phi^a\phi^b\rightarrow m^2_{ef}\left(\delta^{e}{}_a-\epsilon^A\theta^A{}^{e}{}_a\right)
\phi^a\big(\delta^{f}{}_b-\epsilon^B\theta^B{}^{f}{}_b\big)\phi^b\\
=\left[m^2_{ab}-\epsilon^A\left(m^2_{eb}\theta^A{}^{e}{}_a+m^2_{ae}\theta^A{}^{e}{}_b\right)\right]
\phi^a\phi^b+\mathcal{O}\left(\epsilon^2\right).
\end{aligned}
\end{equation}
Since the scalar mass term must be gauge invariant
\begin{equation}
\epsilon^A\big(m^2_{eb}\theta^A{}^{e}{}_a+m^2_{ae}\theta^A{}^{e}{}_b\big)\phi^a\phi^b+\mathcal{O}\left(\epsilon^2\right)= 0,
\end{equation}
it follows that
\begin{equation}
m^2_{eb}\theta^A{}^{e}{}_a+m^2_{ae}\theta^A{}^{e}{}_b=0.
\end{equation}
The same argument can be used to derive the general identity which holds for all the couplings and $W$-coefficients of the operators
\begin{equation}\label{eq:gauge_id}
\begin{aligned}
&if^{B E A_1} W^{(N) B A_2...A_k}_{a_1...a_m}\,+\,\ldots\,+\,if^{B E A_k}W^{(N) A_1...B}_{a_1...a_m}\\ 
&+\,\theta^E_{b a_1} W^{(N) A_1...A_k}_{b a_2...a_m}\,+\,\ldots\,+\,\theta^E_{b a_m}W^{(N) A_1...A_k}_{a_1...b}=0.
\end{aligned}
\end{equation}
As the RGEs for the couplings and $W$-coefficients computed with the
background field method are gauge invariant, the above identity can be
used to bring them to the simplest form.

\section{Passing to the on-shell basis}
\label{sec:On-shell}

The EOMs for the gauge and scalar fields
\begin{align}\label{eq:EOM}
(D_{\mu}F^{\mu \nu})^A&= - i \theta^A_{ab}\phi_b (D^{\nu}\phi)_a+ \mathcal{O}(\frac{1}{\Lambda}),\\
(D_{\mu}D^{\mu}\phi)_a&=-m^2_{ab}\phi_b-\frac{1}{3!}\lambda_{abcd}\phi_b \phi_c \phi_d+\mathcal{O}(\frac{1}{\Lambda}),
\end{align}
yield linear relations
in the off-shell basis. Using these relations, three operators can be
identified as vanishing by the EOM. However, the choice of the
final sub-system that does not vanish on-shell is a matter of
convention. In the present paper, we choose
\begin{equation}\label{eq:Basis_on_shell}
\{\widetilde{Q}_1,\widetilde{Q}^{(X)}_2,\widetilde{Q}_5,\widetilde{Q}_6,\widetilde{Q}_8,\widetilde{Q}_9\},
\end{equation}
where we use the name $Q^{(X)}_2$ instead on $Q_2$, as its
$W$-coefficient (denoted below by $XW^{(2)}$) has more
symmetries in the on-shell basis than it does off-shell,
namely $XW^{(2)}_{abcd} = XW^{(2)}_{cdab}$ and $XW^{(2)}_{(abcd)} =
0$.

The RGEs derived in the off-shell basis must be transformed to
the on-shell one. After passing to the on-shell basis, the RGEs
of the operators that do not vanish under EOM form a closed
subsystem that depends neither on the EOM-vanishing operator
$W$-coefficients, nor on the gauge-fixing parameter.

Once the system of on-shell operators is fixed, the
transformation to the on-shell basis is completely determined. For
an illustrative purpose, we demonstrate our derivation
of the transformation rules for the operator $Q_5$ which in
the off-shell basis is related via the EOM to the operators
$Q_4$ and $Q_7$. Before passing to the on-shell
basis~\eqref{eq:Basis_on_shell} the latter two operators must
be redefined in such a way that their redefined versions vanish
by the EOM. We begin with redefining $Q_7$
\begin{equation}\label{eq:Q7_os}
\widetilde{Q}_{7} := Q_7+\frac{1}{2} Q_4^{\prime} +\frac{1}{4} Q_5^{\prime},
\end{equation}
where
\begin{align}
Q_4^\prime    &:= i W^{(7)AC} \theta^C_{ab}(D^{\mu}\phi)_a(D^{\nu}\phi)_b F^A_{\mu \nu},&
Q_5^{\prime}  &:= \frac{1}{4}  \left( \sum_{\rm perms}  W^{(7)AC} \theta^C_{ac}\theta^B_{bc} \right)
                  \phi_a \phi_b F^A_{\mu \nu} F^{B\,\mu\nu},
\nonumber\\
\end{align}
and the sum in $Q_5^{\prime}$ runs over all permutations of uncontracted scalar and group indices.
Next, $Q_4^\prime$ and $Q_5^\prime$ are absorbed into $Q_4$ and $Q_5$:
\begin{align}\label{eq:W4_os}
\overline{W}^{(4)A}_{ab}  &:= W^{(4)A}_{ab}-i W^{(7)AC}\theta^C_{ab},&
\overline{W}^{(5)AB}_{ab} &:= W^{(5)AB}_{ab}- \frac{1}{4} W^{(7)AC} \theta^C_{ac}\theta^B_{bc}\,.
\end{align}
The fully redefined operator $Q_4$, denoted below by
$\widetilde{Q_{4}}$, must identically vanish on-shell. This is
achieved with the help of yet another redefinition:
\begin{equation}\label{eq:Q4_os}
\widetilde{Q_{4}} := Q_4+\tfrac{1}{4} Q^{\prime\prime}_5+(\ldots),
\end{equation}
where
\begin{equation}
Q^{\prime\prime}_5:=
\tfrac{i}{4}\left(\sum_{\rm perms}\overline{W}^{(4)A}_{ac}\theta^B_{cb}\right)\phi_a \phi_b F^A_{\mu \nu} F^{B\,\mu\nu},
\end{equation}
and dots denote terms that are irrelevant for the $Q_5$ transformation.

The operators $Q_4$ and $Q_7$ are the only ones that matter
for the redefined $W$-coefficient of $Q_5$. Therefore, the
redefinitions~\eqref{eq:Q7_os} and~\eqref{eq:Q4_os} followed by
the necessary redefinitions of the $W$-coefficients are sufficient to
get an on-shell expression for the $W$-coefficient of
$\widetilde{Q}_5$, namely
\begin{align}\label{eq:W5_os}
\widetilde{W}^{(5)AB}_{ab} &= \overline{W}^{(5)AB}_{ab}
+\frac{i}{4} \sum_{\rm perms} \overline{W}^{(4)A}_{ac}\theta^B_{bc}
= W^{(5)AB}_{ab} +\frac{i}{4} \sum_{\rm perms} W^{(4)A}_{ac}\theta^B_{bc}\,.
\end{align}
The RGE for $\widetilde{W}^{(5)}$ is obtained simply by applying $\mu\frac{d}{d\mu}$ to both sides of the above equation
\begin{eqnarray}
\mu\frac{d\widetilde{W}^{(5)AB}_{ab}}{d\mu} &=& \mu\frac{dW^{(5)AB}_{ab}}{d\mu}
+ \frac{i}{4} \sum_{\rm perms}\left( \mu\frac{d W^{(4)A}_{ac}}{d\mu}\theta^B_{bc}
+ W^{(4)A}_{ac} \theta^{\underline{B}}_{bc} \gamma_{\underline{B}}  \right),
\end{eqnarray}
with
\begin{equation}
\gamma_{B}=\frac{1}{48\pi^2} \left[ -11 C_2(G_B)
+ \frac{1}{2}S_2(S_B) \right].
\end{equation}

All our transformation rules are obtained by 
applying an analogous procedure for the remaining 
EOM-vanishing operators, i.e.\ $Q_3$ and the unphysical parts of
$Q_2$. The full list of the transformation rules for the
operators~\eqref{eq:Basis_on_shell} is given in appendix~\ref{app:trans_rules}.

\section{Renormalization group equations for the bosonic operators}
\label{sec:RGE}

\subsection{Methodology}
\label{ssec:methodology}

To renormalize the $W$-coefficients of dimension-six bosonic
operators in the off-shell basis, we proceeded by computing all
the one-loop, 1-particle-irreducible diagrams with the external fields
matching the field content of the renormalized operators.\footnote{
Although such computation can be completed in the general $R_\xi$
gauge with an arbitrary value of the gauge parameter,
we have chosen the Feynman-'t~Hooft gauge ($\xi=1$)
to reduce the problem complexity.}
The task was handled using standard symbolic computation
tools. First, all the necessary Feynman rules were derived from the
Lagrangian~\eqref{eq:Lagrangian} with the Mathematica package
\texttt{FeynRules}~\cite{Alloul:2013bka}. Next, all diagrams
potentially relevant to the problems were generated and translated to
Feynman integrands using the dedicated package
\texttt{FeynArts}~\cite{Hahn:2000kx}. Divergent parts were extracted
using our code, whose performance was validated against the
\texttt{FeynCalc} package~\cite{Shtabovenko:2020gxv}. The off-shell
results were simplified using yet another package, namely
\texttt{xTensor}~\cite{Martin-Garcia:2007bqa,xAct:2002xx} that is
dedicated to symbolic tensor algebra.

The off-shell results were subsequently translated to the on-shell
basis and reduced to their simplest form with the help of the
gauge identities~\eqref{eq:gauge_id}, using a combination of custom
codes and manual computations. Although conceptually
straightforward, this step proved to be the most challenging part of
the calculation. To address it, we followed the procedure
outlined in Ref.~\cite{Poole:2019kcm}. First, for each
$W$-coefficient, we completed a classification of the symmetry-allowed
expressions that could potentially contribute to its RGE at a given
order in the couplings. Next, we systematically derived gauge
relations among the elements of this set. These relations were used to
find a minimal system in which any valid combination of
$W$-coefficients and couplings can be unambiguously represented. While
constructing this minimal system of irreducible structures, we
prioritized contractions with explicit group invariants, such as group
or representation Casimir operators, over more cumbersome
constructions involving multiple scalar generators and structure
constants. Finally, the on-shell results were expressed in terms of
the elements of the minimal system.

\subsection{On-shell results}
\label{ssec:results}

In this section, we present our main results, namely the on-shell RGEs
for all $W$-coefficients of the physical dimension-six bosonic
operators. Since only the on-shell results are considered here, 
we suppress tildes over the $W$ and $Q$ symbols to make the
notation simpler. Moreover, we introduce rescaled anomalous dimensions
\begin{align}
&\eta_B := (4\pi)^2\gamma_{B}=\frac{1}{3} \left[ -11 C_2(G_B)
+\frac{1}{2}S_2(S_B) \right],\\[2mm]
&(\eta_\phi)_{ab}:= (4\pi)^2 (\gamma_\phi)_{ab}= -2\,  C_2(S)_{ab},
\end{align}
to absorb the $(4\pi)^2$ factors stemming from the 
corresponding counterterms. The matrix $C_2(S)_{ab}:=
\theta^A_{ac} \theta^A_{cb}$ that appears in
$(\eta_\phi)_{ab}$ collects the Casimir invariants of the
scalar representations.

The most involved RGE is encountered for the operator $Q_1$ that
contains the scalar fields only. We find
\vspace*{0.3cm}
\begin{flalign}
\mu\frac{dW^{(1)}_{abcdef}}{d\mu}\,=\,\frac{1}{(4\pi)^2}\big(&-A^{(1)} + A^{(2)} + A^{(3)} 
+ 6\, A^{(4)} +3\, A^{(5)}+\tfrac{3}{2}\, A^{(6)}+2\,A^{(7)}-\tfrac{1}{2}\,A^{(8)}&&\\\nonumber
&-12 \,A^{(9)}+6\,A^{(10)}\big)_{abcdef},
\end{flalign}
where the tensors $A^{(1)}$-$A^{(10)}$ are contractions of the
$W$-coefficients, the gauge-group generators, and the scalar
quartic coupling matrices $\lambda_{ijkl}$. Explicitly:
\begin{equation}
\begin{aligned}[c]
& A^{(1)}_{abcdef}=\tfrac{1}{120}\sum_{\text{perms}}W^{(1)}_{abcdeg} C_2(S)_{gf}, \\
&A^{(3)}_{abcdef}=\tfrac{1}{48}\sum_{\text{perms}}W^{(1)}_{abcdgh} \lambda_{ghef}, \\
& A^{(5)}_{abcdef}=\tfrac{1}{6}\sum_{\text{perms}}XW^{(2)}_{ahib} \theta^A_{hc} \theta^A_{ig} \lambda_{gdef},\\
& A^{(7)}_{abcdef}=\tfrac{1}{16}\sum_{\text{perms}}XW^{(2)}_{abgh} \lambda_{gicd}\lambda_{hief},\\ 
& A^{(9)}_{abcdef}=\tfrac{1}{4}\sum_{\text{perms}}W^{(5)AB}_{ab} \theta^C_{cg} \theta^A_{gd} \theta^C_{eh} \theta^B_{hf},
\end{aligned}
\qquad
\begin{aligned}[c]
& A^{(2)}_{abcdef}=\tfrac{1}{120}\sum_{\text{perms}}W^{(1)}_{abcdeg} (\eta_\phi)_{gf}, \\
& A^{(4)}_{abcdef}=\tfrac{1}{4}\sum_{\text{perms}}XW^{(2)}_{abgh}\theta^A_{gc}\theta^B_{hd}\theta^A_{ei}\theta^B_{if}, \\
& A^{(6)}_{abcdef}=\tfrac{1}{6}\sum_{\text{perms}}XW^{(2)}_{ahgb} C_2(S)_{hc} \lambda_{gdef},\\
& A^{(8)}_{abcdef}=\tfrac{1}{12}\sum_{\text{perms}}XW^{(2)}_{abhi} \lambda_{ghic}\lambda_{gdef},\\
& A^{(10)}_{abcdef}=\tfrac{1}{12}\sum_{\text{perms}}W^{(5)AB}_{ab} \theta^A_{ch} \theta^B_{hg} \lambda_{gdef}.
\end{aligned}
\end{equation}
Here, the sums in the $A$-structures are taken
over such permutations of uncontracted indices under which the $W$-coefficient
on the l.h.s.\ of the considered RGE is invariant. Each
$A$-structure is normalized to the number of index
permutations that leave each term in the sum invariant. For
instance, in the structure $A^{(1)}_{abcdef}$, the summation
runs over all possible permutations of scalar indices, as
$W^{(1)}$ is fully symmetric. Nevertheless, the contraction
$W^{(1)}_{abcdeg} C_2(S)_{gf}$ in $A^{(1)}$ is already fully
symmetric under permutations of its first five indices, which yields
the normalization factor equal to $1/5!=1/120$.

The RGEs for the on-shell operators describing the interaction of
scalar and gauge fields are found to be as follows:\footnote{
Both sides of the RGE for $XW^{(2)}_{abcd}$ in
Eq.~\eqref{eq:XW2_RGE} are contracted with an arbitrary
tensor $TW^{(2)}_{abcd}$ that is symmetric in the first and
last pairs of its indices, to enforce the original
symmetry of $W^{(2)}$. Such a notation allows for suppressing redundant
terms that vanish when $XW^{(2)}$ is contracted with the fields, and
reduces the length of its RGE.}
\begin{flalign} \label{eq:XW2_RGE}
\mu\frac{dXW^{(2)}_{abcd}}{d\mu}TW^{(2)}_{abcd}=\frac{1}{(4\pi)^2}\big(&- 2\, B^{(1)} - 2 \, B^{(2)} +2\, B^{(3)}
-\tfrac{4}{3}\,B^{(4)}+\tfrac{1}{3}\,B^{(5)} - \tfrac{2}{3} \,B^{(6)}&&\\ \nonumber
&+ \tfrac{2}{3} \, B^{(7)}+ \tfrac{2}{3} \, B^{(8)})_{abcd}TW^{(2)}_{abcd}, 
\end{flalign}
\vspace{-0.85cm}
\begin{flalign}
&\mu\frac{dW^{(5)AB}_{ab}}{d\mu}=\frac{1}{(4\pi)^2}(2\, C^{(1)} -\, C^{(2)} -3\, C^{(3)} + C^{(4)}+2\,C^{(5)}+
C^{(6)}+\tfrac{1}{2}C^{(7)}-3\,C^{(8)})^{AB}_{ab},&&\\[2mm]
&\mu\frac{dW^{(6)AB}_{ab}}{d\mu}=\frac{1}{(4\pi)^2}(2\, \widetilde{C}^{(1)} -\, \widetilde{C}^{(2)} -3\, \widetilde{C}^{(3)}
+ \widetilde{C}^{(4)}+2\,\widetilde{C}^{(5)}+ \widetilde{C}^{(6)}+\tfrac{1}{2}\widetilde{C}^{(7)}-3\,\widetilde{C}^{(8)})^{AB}_{ab},
\end{flalign}
with the $B$-structures defined by
\begin{equation}
\begin{aligned}[c]
& B^{(1)}_{abcd}=\tfrac{1}{2}\sum_{\text{perms}} XW^{(2)}_{edfc} \theta^A_{ea} \theta^A_{fb}, \\
& B^{(3)}_{abcd}=\tfrac{1}{2}\sum_{\text{perms}} XW^{(2)}_{abef} \theta^A_{ec} \theta^A_{fd}, \\
& B^{(5)}_{abcd}=\sum_{\text{perms}} XW^{(2)}_{abed} C_2(S)_{ec}, \\
& B^{(7)}_{abcd}=\tfrac{1}{2}\sum_{\text{perms}} XW^{(2)}_{cdef} \lambda_{abef},
\end{aligned}
\qquad\qquad
\begin{aligned}[c]
& B^{(2)}_{abcd}=\tfrac{1}{2}\sum_{\text{perms}} XW^{(2)}_{aefd} \theta^A_{eb} \theta^A_{fc}, \\[1mm]
& B^{(4)}_{abcd}= XW^{(2)}_{aefd}\theta^A_{ef}\theta^A_{bc}, \\[2mm]
& B^{(6)}_{abcd}=\tfrac{1}{2}\sum_{\text{perms}} XW^{(2)}_{cedf} \lambda_{abef}, \\
& B^{(8)}_{abcd}=\sum_{\text{perms}} XW^{(2)}_{abce} (\eta_\phi)_{ed}.
\end{aligned}
\end{equation}
The sums appearing in the $B$-structures run over
permutations of the first and third indices, as well as
the second and last indices. They also assume
antisymmetrization under the exchange of the second and third index,
which is necessary to remove the symmetric part of
$XW^{(2)}$. Altogether, these (anti)symmetrizations
correspond to the projection of the off-shell RGE for
$W^{(2)}$ onto the $XW^{(2)}$ subspace.

The $C$-structures in the RGE for $W^{(5)}$ read
\begin{equation} 
\begin{aligned}[c]
& C^{(1)AB}_{ab}=\sum_{\text{perms}} W^{(5)AC}_{ac} \theta^C_{cd} \theta^B_{db}, \\
& C^{(3)AB}_{ab}=i \sum_{\text{perms}} f^{ACD} W^{(5)BC}_{ac} \theta^D_{cb}, \\[1mm]
& C^{(5)AB}_{ab}= \eta_{\underline{B}}  W^{(5)A\underline{B}}_{ab}, \\[2mm]
& C^{(7)AB}_{ab}=\sum_{\text{perms}} W^{(8)ACD} f^{ECD} \theta^B_{ac} \theta^E_{cb},\\
\end{aligned}
\qquad
\begin{aligned}[c]
& C^{(2)AB}_{ab}=\tfrac{1}{2}\sum_{\text{perms}} W^{(5)AB}_{ac} C_2(S)_{cb}, \\[1mm]
& C^{(4)AB}_{ab}= W^{(5)AB}_{cd} \lambda_{abcd},  \\[2mm]
& C^{(6)AB}_{ab}=\tfrac{1}{2}\sum_{\text{perms}} W^{(5)AB}_{bc} (\eta_\phi)_{ca}, \\
& C^{(8)AB}_{ab}=\sum_{\text{perms}} W^{(8)AED} f^{BEC} \theta^C_{ac} \theta^D_{cb},
\end{aligned}
\end{equation}
where the sums run over permutations of the uncontracted scalar and gauge
indices. The $\widetilde{C}$-structures in the RGE for
$W^{(6)}$ are recovered from the above $C$-structures via simple
substitutions $W^{(5)AB}_{ab}\rightarrow W^{(6)AB}_{ab}$ and
$W^{(8)ABC}\rightarrow W^{(9)ABC}$.

Finally, the one-loop RGEs for the $W$-coefficients of $Q_8$ and $Q_9$ are 
\begin{eqnarray}
\mu \frac{dW^{(8)ABC}}{d\mu} &= \frac{1}{(4\pi)^2}
[ 12\,C_2(G_{\underline{B}})+3\,\eta_{\underline{B}} ]\,W^{(8)A\underline{B}C},\label{eq:w8}\\
\mu \frac{dW^{(9)ABC}}{d\mu} &= \frac{1}{(4\pi)^2}
[ 12\,C_2(G_{\underline{B}})+3\,\eta_{\underline{B}}]\,W^{(9)A\underline{B}C}\label{eq:w9}.
\end{eqnarray}
Although these two equations have been well-known for over 30 years~\cite{Braaten:1990gq,Braaten:1990zt},
we recall them here for completeness. 

As expected, in the on-shell basis, the r.h.s.\ of the RGEs are free
of the redefined off-shell coefficients $W^{(3)}$, $W^{(4)}$
$W^{(7)}$. The on-shell-redundant parts of $W^{(2)}$ cancel out, too,
which constitutes a strong consistency check of our results. The
apparent symmetry between the RGEs for the operators with the regular
and dual field strength tensors can be traced to the
chirality-flip symmetry that exchanges the left-handed and
right-handed polarizations of the spin-1 gauge bosons. When
deriving our results, we have not imposed any symmetry
of this kind.

To further test our results, we have used them to
independently derive the one-loop RGEs in the SMEFT-like extension of
the SM with gauge and scalar fields only. The substitutions of the
necessary group invariants was accomplished with the help of a
simple code written in Mathematica, available
at~\cite{Repo:2025in}. The equations obtained after substitutions 
do match the bosonic part of the full one-loop RGEs in the
SMEFT~\cite{Jenkins:2013zja,Jenkins:2013wua,Alonso:2013hga}. 
This is yet another successful test of our general RGEs.

\section{Summary}
\label{sec:summ}

In this work, we derived generalized 1-loop renormalization
group equations for the bosonic operators of dimension six. The
renormalization of the operator coefficients was initially performed
in the off-shell basis, which mixes the operators that vanish under
EOM with non-vanishing ones. Next, the results were 
converted to the on-shell basis where only the physical
operators are present. Only in the latter basis our results are
convention-independent and can be applied to reproduce RGEs for
specific models with particular gauge groups and matter
contents.

Our final results were validated with two independent consistency
checks. First, through the whole computation we carefully traced all
the coefficients of the EOM-vanishing operators, assuring their
cancellation at the final stage. Second, we identified symmetries
relating some operator anomalous dimensions that follow from the
structure of the general EFT, and \emph{a posteriori} checked
that our RGEs in the final form do respect them.

The results of this work tremendously simplify derivation of one-loop
RGEs for dimension-six operator coefficients in a wide class of
bosonic EFTs. For validation purposes, we 
re-derived all the one-loop equations in the simplified version of
SMEFT with no fermionic fields. The number of effective operators in
this model is large~\cite{Grzadkowski:2010es}, which significantly
complicates most computations. In contrast, making appropriate
substitutions to our results required preparing only a
simple script, which generated the results within a few
seconds. The obtained equations fully agree with the corresponding
part of SMEFT
RGEs~\cite{Jenkins:2013zja,Jenkins:2013wua,Alonso:2013hga}, providing
yet another useful verification of our findings.

This paper is part of a broader project that aims at deriving
the RGEs for a wide class of EFTs. In the next
stage~\cite{Mieszkalski:2021xx}, we are going to present complete
one-loop results for dimension-six operators in the EFTs that, 
apart from the gauge and scalar fields, include also 
the spin-$\frac12$ fermions. Such results could be used
to independently derive the full one-loop SMEFT RGEs, and would
represent an important step towards determining two-loop
RGEs in a generic EFT. The two-loop results would enable
us to obtain the corresponding SMEFT RGEs via straightforward
substitutions. The complete, one-loop results may be also
useful for theories with light, weakly coupled particles beyond the
Standard Model. When a high-energy theory unifies such hypothetical
light particles with the known ones, the  SMEFT operators 
might inadequately describe the physical reality. However, most of
such scenarios can still be effectively captured using the
generic EFT formalism developed in this work.

\vspace{5mm}
\noindent\textit{Note added}:
When this article was being finalized, a paper~\cite{Fonseca:2025zjb}
with significant overlap appeared on the arXiv. Our main on-shell results in 
section~\ref{sec:RGE} can be compared to Eqs.~(107)-(112) of that paper\footnote{
The quoted equation numbering corresponds to the original (v1) arXiv version of Ref.~\cite{Fonseca:2025zjb}.}
in the limit $a^{(5)} \to 0$ and vanishing fermion-loop contributions
on their side, as neither the fermions nor dimension-five operators
have been considered in our paper. In this limit, we find perfect
agreement between their Eqs.~(107)-(108) and our results for $W^{(8)}$
and $W^{(9)}$ in Eqs.~(\ref{eq:w8})-(\ref{eq:w9}), as well as between
their Eqs.~(110)-(111) and our results for $W^{(5)}$ and $W^{(6)}$ in
section~\ref{sec:RGE}. Actually, Eqs.~(110)-(111) of
Ref.~\cite{Fonseca:2025zjb} can be further simplified to reveal the
apparent symmetry between the RGEs for $W^{(5)}$ and $W^{(6)}$ in our
results. The remaining two cases, namely $W^{(1)}$ and $W^{(2)}$, are
much more involved. As far as $W^{(1)}$ is concerned, we find a sign
difference in the term in Eq.~(112) of Ref.~\cite{Fonseca:2025zjb}
that reads $-\frac18 g^2 \theta^A_{ha} \theta^A_{gb} \left(
a_\phi^{(6)} \right)_{cdefhg}$.  Only with our sign for this term, we
can reproduce the RGE of Ref.~\cite{Alonso:2013hga} for the single
SMEFT operator that contains the Higgs boson field raised to the sixth
power. Furthermore, our RGE for $W^{(1)}$ contains an additional
term proportional to the contraction of $XW^{(2)}$ with the Casimir of
the scalar representation, which does not reduce to the combination of
the tensor structures given in Eq.~(112) of
Ref.~\cite{Fonseca:2025zjb}.  The remaining terms in the RGE
for $W^{(1)}$ are found to be equivalent. As far as the RGE
for $W^{(2)}$ is concerned, our comparison has not yet been
completed.

\section*{Acknowledgments}

We would like to acknowledge Patryk Mieszkalski for his contributions
at the early stage of this project. We also thank Anders Thomsen for
useful discussions about the reduction of complex algebraic
structures, and Aneesh Manohar for sharing his ideas regarding
renormalization of the operators with dual gauge tensors. We are
grateful to the authors of Ref.~\cite{Fonseca:2025zjb} for discussions
in 2022 concerning the on-shell projection of the operator $Q_2$ in
Eq.~(\ref{eq:Basis_off_shell}).  Finally, we acknowledge the
willingness of Jason Aebisher and Nud\v{z}eim Selimovi\'c to join our
project starting from 2022, even though they eventually continued on
their own~\cite{Aebischer:2025zxg}, and none of their results
had been known to us prior to the original arXiv submission
of the present paper. This research was supported by the National
Science Center, Poland, under the research grants 2020/37/B/ST2/02746
and 2023/49/B/ST2/00856 in the case of MM, as well as
2020/38/E/ST2/00243 in the case of IN.

\appendix

\section{Transformation rules to the on-shell basis}
\label{app:trans_rules}

In this appendix, we list the transformation rules of
the off-shell $W$-coefficients from~\eqref{eq:Basis_off_shell} to the
on-shell basis~\eqref{eq:Basis_on_shell}. The coefficients of
$\widetilde{Q}_1$, $\widetilde{Q}^{(X)}_2$ and $\widetilde{Q}_5$
are~\cite{Mieszkalski:2021xx}
\begin{align}
    &\widetilde{W}^{(1)}_{abcdef}:= W^{(1)}_{abcdef} +\sum_{\sigma(a \ldots f)} \Big[\frac{1}{24}\,(RW^{(2)}
    +\frac{1}{3}SW^{(2)})_{gabc}\lambda_{gdef}+\frac{1}{72}W^{(3)}_{gh}\lambda_{gabc}\lambda_{hdef}\Big],\\[2mm]
    &X\widetilde{W}^{(2)}_{abcd}:= XW^{(2)}_{abcd}
    - \frac{i}{2}\sum_{\sigma(ab)\times\sigma(cd)}\overline{W}^{(4)A}_{ad}\theta^A_{bc},\\[2mm]
    &\widetilde{W}^{(5)AB}_{ab} =  W^{(5)AB}_{ab}
    +\frac{i}{4} \sum_{\sigma(ab)\times\sigma(AB)} W^{(4)A}_{ac}\theta^B_{bc}\,,
\end{align}
where the $\sigma$ symbols in the sums denote
summations over all the indicated index
permutations. The coefficient $\overline{W}^{(4)}$ was defined
in Eq.~\eqref{eq:W4_os}. Our intermediate-step RGEs in the
off-shell basis that are already partly written in terms of the
on-shell basis coefficients read
\begin{align}
    &\mu\frac{d\widetilde{W}^{(1)}_{abcdef}}{d\mu}=\mu\frac{d W^{(1)}_{abcdef}}{d\mu}+\sum_{\sigma(a\ldots f)}
    \Big[\frac{1}{24}\,\Big(\mu\frac{d RW^{(2)}}{d\mu}+\frac{1}{3}\mu\frac{d SW^{(2)}}{d\mu}\Big)_{gabc}\lambda_{gdef}\\ \nonumber
    &+\frac{1}{24}\Big(RW^{(2)}+\frac{1}{3}SW^{(2)}\Big)_{gabc}(\beta_{\lambda})_{gdef}
     +\frac{1}{72}\mu\frac{d W^{(3)}_{gh}}{d\mu} \lambda_{gabc}\lambda_{hdef}\\ \nonumber
    &+\frac{1}{36} W^{(3)}_{gh}(\beta_{\lambda})_{gabc}\lambda_{hdef}\Big],\\[2mm]
    &\mu\frac{dX\widetilde{W}^{(2)}_{abcd}}{d\mu}=\mu\frac{d XW^{(2)}_{abcd}}{d\mu}
    -\frac{i}{2}\sum_{\sigma(ab)\times \sigma(cd)}\Big(\mu\frac{d\widetilde{W}^{(4)A}_{ad}}{d\mu}\theta^A{}_{bc}
    +\beta_g \widetilde{W}^{(4)A}_{ad}\frac{\partial \theta^A_{bc}}{\partial g}\Big),\\[2mm]
    &\mu\frac{d \widetilde{W}^{(5)AB}_{ab}}{d\mu} = \mu\frac{dW^{(5)AB}_{ab}}{d\mu}
    + \frac{i}{4} \sum_{\sigma(ab)\times \sigma(AB)}\left( \mu\frac{d W^{(4)A}_{ac}}{d\mu}\theta^B_{bc}
    + W^{(4)A}_{ac} \theta^{\underline{B}}_{bc} \gamma_{\underline{B}}  \right),
\end{align}
where $RW^{(2)}:=\frac{1}{2}\big(W^{(2)}_{abcd}-W^{(2)}_{cdab}\big)$ 
and $SW^{(2)}:=W^{(2)}_{(abcd)}$  are the $W$-coefficients of the EOM-reducible parts of $Q_2$.
The symbol $\beta_\lambda$ denotes the one-loop beta function for the
scalar quartic coupling that reads
\begin{equation}
(\beta_{\lambda})_{abcd}=\frac{1}{(4\pi)^2}[\Lambda^2+3\,A-3\,\Lambda^{S}]_{abcd},
\end{equation}
with
\begin{equation}
\begin{aligned}
& \Lambda^2_{abcd} = \tfrac{1}{8}\sum_{\sigma(abcd)}\lambda_{abef}\lambda_{cdef},\qquad A_{abcd}=
\tfrac{1}{8}\sum_{\sigma(abcd)}\{\theta^A,\theta^B\}_{bc}\{\theta^A,\theta^B\}_{ad},\\
& \Lambda^S_{abcd}=\tfrac{1}{6}\sum_{\sigma(abcd)}C_2(S)_{ae}\lambda_{ebcd}.
\end{aligned}
\end{equation}
The remaining three coefficients ($W^{(6)}$, $W^{(8)}$ and $W^{(9)}$) transform trivially to the on-shell basis.

\bibliography{main.bib}
\bibliographystyle{JHEP}
\end{document}